\pdfoutput=1

\documentclass[aps,prd,twocolumn,preprintnumbers,superscriptaddress,nofootinbib,floatfix]{revtex4-1}

\usepackage{bm}

\usepackage[hyperfootnotes=false]{hyperref}
\usepackage{url}

\usepackage{color}
\usepackage[usenames,dvipsnames,svgnames,table]{xcolor}

\usepackage{amsmath}
\usepackage{amssymb}
\usepackage{graphicx}
\usepackage{slashed}
\usepackage{soul}

\def\beq{\begin{equation}}
\def\eeq{\end{equation}}
\def\bea{\begin{eqnarray}}
\def\eea{\end{eqnarray}}
\def\ben{\begin{enumerate}}
\def\een{\end{enumerate}}

\def\lsim{\mathrel{\raise.3ex\hbox{$<$\kern-.75em\lower1ex\hbox{$\sim$}}}}
\def\gsim{\mathrel{\raise.3ex\hbox{$>$\kern-.75em\lower1ex\hbox{$\sim$}}}}
\def\ifmath#1{\relax\ifmmode #1\else $#1$\fi}

\newcommand{\ccHal}{NaCl}
\newcommand{\ccEps}{MgSO$_4\!\cdot\!7($H$_2$O)}
\newcommand{\ccOli}{Mg$_{1.6}$Fe$^{2+}_{0.4}$(SiO$_4$)}
\newcommand{\ccNic}{NiCl$_2\!\cdot\!6($H$_2$O)}

\begin{document}

\title{Searching for Dark Matter with Paleo-Detectors} 

\newcommand{\OKC}{\affiliation{The Oskar Klein Centre for Cosmoparticle Physics, Department of Physics, Stockholm University, Alba Nova, 10691 Stockholm, Sweden}}
\newcommand{\Nordita}{\affiliation{Nordita, KTH Royal Institute of Technology and Stockholm University, Roslagstullsbacken 23, 10691 Stockholm, Sweden}}
\newcommand{\LCTP}{\affiliation{Leinweber Center for Theoretical Physics, University of Michigan, Ann Arbor, MI 48109, USA}}
\newcommand{\NCBJ}{\affiliation{National Centre for Nuclear Research, 05-400 Otwock, \'{S}wierk, Poland}}
\newcommand{\UTexas}{\affiliation{Department of Physics, University of Texas, Austin, TX 78712, USA}}
\newcommand{\SITP}{\affiliation{Stanford Institute for Theoretical Physics, Department of Physics, Stanford University, Stanford, CA 94305, USA}}

\author{Sebastian~Baum}
\email{sbaum@stanford.edu}
\OKC
\Nordita
\SITP

\author{Andrzej~K.~Drukier}
\email{adrukier@gmail.com}
\OKC

\author{Katherine~Freese}
\email{ktfreese@umich.edu}
\OKC
\Nordita
\LCTP
\UTexas

\author{Maciej~G\'{o}rski}
\email{maciej.gorski@ncbj.gov.pl}
\NCBJ

\author{Patrick~Stengel}
\email{patrick.stengel@fysik.su.se}
\OKC

\preprint{NORDITA-2018-043}
\preprint{LCTP-18-15}

\begin{abstract}
A large experimental program is underway to extend the sensitivity of direct detection experiments, searching for interaction of Dark Matter with nuclei, down to the {\it neutrino floor}. However, such experiments are becoming increasingly difficult and costly due to the large target masses and exquisite background rejection needed for the necessary improvements in sensitivity. We investigate an alternative approach to the detection of Dark Matter--nucleon interactions: Searching for the persistent traces left by Dark Matter scattering in ancient minerals obtained from much deeper than current underground laboratories. We estimate the sensitivity of paleo-detectors, which extends far beyond current upper limits for a wide range of Dark Matter masses. The sensitivity of our proposal also far exceeds the upper limits set by Snowden-Ifft {\it et al.} more than three decades ago using ancient Mica in an approach similar to paleo-detectors.
\end{abstract}

\maketitle

\section{Introduction} The gravitational effects of Dark Matter (DM) are evident at length scales ranging from the smallest galaxies to the largest observable scales of our Universe. However, despite much experimental effort, the nature of DM is as yet unknown.

{\it Weakly Interacting Massive Particles} (WIMPs) are well motivated DM candidates. A large ongoing experimental program exists to search for WIMP induced nuclear recoil events in {\it direct detection} experiments~\cite{Goodman:1984dc,Drukier:1983gj}; signatures include annual~\cite{Drukier:1986tm,Spergel:1987kx,Freese:2012xd} and diurnal modulation~\cite{Collar:1992qc}. The only experiment with positive results is DAMA, which reports a $12\,\sigma$ measurement of an annually modulated signal compatible with WIMP DM~\cite{Bernabei:2008yi,Bernabei:2013xsa,Bernabei:2018yyw}. However, this result is in tension with null results from other direct detection experiments which have set stringent limits on the WIMP-nucleus interacting strength~\cite{Angloher:2015ewa,Amole:2017dex,Akerib:2017kat,Aprile:2017iyp,Agnese:2017jvy,Cui:2017nnn,Petricca:2017zdp,Aprile:2018dbl,Agnes:2018ves,Abdelhameed:2019hmk}.

In the foreseeable future, direct detection experiments are expected to push into two different directions: Large scale detectors using e.g. liquid noble gas targets, aim to obtain exposures (defined as the product of target mass and integration time) of $\varepsilon = \mathcal{O}(10)\,$t\,yr in the next few years~\cite{Aprile:2015uzo,Mount:2017qzi,Amaudruz:2017ibl,Aalseth:2017fik} and $\varepsilon = \mathcal{O}(100)\,$t\,yr in the next decades~\cite{Aalseth:2017fik,Aalbers:2016jon}. For WIMPs with $m_\chi \lesssim 15\,$GeV, the challenge is achieving low recoil energy thresholds rather than large detector masses; solid state detectors are envisaged to collect $\varepsilon = \mathcal{O}(10)\,$kg\,yr with recoil energy thresholds of $\mathcal{O}(100)\,$eV~\cite{Angloher:2015eza,Agnese:2016cpb}. Recently, nm-scale detectors~\cite{Drukier:2012hj,Drukier:2017piu} and concepts using molecular biology~\cite{Lopez:2014oea,Drukier:2014rea} have been proposed to search for low-mass WIMPs. Directional detectors~\cite{Spergel:1987kx} are being developed with the capability of determining the direction of the incoming WIMPs~\cite{Daw:2011wq,Battat:2014van,Riffard:2013psa,Santos:2013hpa,Monroe:2012qma,Leyton:2016nit,Miuchi:2010hn,Nakamura:2015iza, Battat:2016xxe}.

In this letter we investigate an alternative approach to search for WIMP-nucleus scattering. Instead of building a dedicated target instrumented to search for nuclear recoils in real time, we propose to search for the traces of WIMP interactions in a variety of ancient minerals.  Many years ago in a similar approach, Refs.~\cite{Price:1983ax,Price:1986ky,SnowdenIfft:1995ke,Engel:1995gw,SnowdenIfft:1997hd,Collar:1994mj} looked at using ancient Mica to detect monopoles and, subsequently, as a DM detector; see also Refs.~\cite{Goto:1958,Goto:1963zz,Fleischer:1970zy,Alvarez:1970zu,Kolm:1971xb,Eberhard:1971re,Ross:1973it,Kovalik:1986zz,Jeon:1995rf,Fleischer:1969mj,Fleischer:1970vm,Ghosh:1990ki,Collar:1999md} for related work. We propose to use a variety of new materials and analysis techniques. Whereas in conventional direct detection experiments the observable is the prompt energy deposition from nuclear recoils in the target material read out via e.g. scintillation, ionization, or phonons, here the observable is the persistent chemical or structural change caused in the material by the recoiling nucleus. Conventional direct detection experiments require high quantum efficiency of the detector to read out the $\mathcal{O}(1)$ number of phonons, photons, or electrons produced by the nuclear recoil. In our case one needs to detect tracks with lengths of $1-1000\,$nm.

To avoid cosmic ray backgrounds, direct detection experiments are typically located in deep underground laboratories. Similarly, we propose to use minerals obtained from large depths for paleo-detectors. While the currently deepest conventional direct detection laboratories offer $\sim 2\,$km of rock overburden, target minerals for paleo-detectors could be obtained from much deeper, e.g. from the cores of deep boreholes used for geological R\&D and oil exploration (see, e.g. Refs.~\cite{Blattlereaar2687,Hirschmann1997,KREMENETSKY198611}). As we will discuss further below, for an overburden larger than $\sim 5\,$km of rock, cosmogenic backgrounds would become virtually negligible. 

Direct detection experiments also suffer backgrounds from radioactive processes in the detector material and its surroundings. To mitigate these, any direct detection apparatus must be built from as radiopure materials as possible. Likewise, for paleo-detectors the most radiopure minerals must be used as target material. Typical minerals formed in the Earth's crust exhibit prohibitively large (order parts per million by weight) contaminations with heavy radioactive trace elements such as $^{238}$U. Thus, we propose to use minerals found in Marine Evaporite (ME) and Ultra Basic Rock (UBR) deposits for paleo-detectors. Such minerals form from sea water and material from the Earth's mantle, respectively, which are much more radiopure than the Earth's crust. In the remainder of this letter, we assume $^{238}$U benchmark concentrations of $C^{238} = 0.01\,$ppb (parts per billion) in weight for MEs and $C^{238} = 0.1\,$ppb for UBRs~\cite{Adams:1959}; see also the Appendix of Ref.~\cite{Baum:2019fqm} for a discussion of $^{238}$U concentrations in natural minerals. For definiteness we use halite (\ccHal), epsomite [\ccEps], olivine [\ccOli], and nickelbischofite [\ccNic] in this work. The former two are examples of minerals found in MEs, while the latter two are examples of minerals found in UBRs.

In the remainder of this letter, we estimate the sensitivity of paleo-detectors, focusing on spin-independent (SI) WIMP-nucleus interactions as an example; we expect our results to generalize to other types of interactions. In Sec.~\ref{sec:Signal} we discuss the track length spectra which would arise from DM in paleo detectors and in Sec.~\ref{sec:Backgrounds} different sources of backgrounds and how to mitigate them. We Sec.~\ref{sec:ReadOut_Sens} we identify promising techniques for the read out of the damage tracks induced by DM and set the stage for projecting the sensitivity of paleo-detectors to DM, which we present in Sec.~\ref{sec:Results}. We reserve Sec.~\ref{sec:Discussion} for a summarizing discussion. A more detailed discussion of our proposal can be found in Ref.~\cite{Drukier:2018pdy}.

Before continuing, let us note that the sensitivity of our proposal far exceeds that of Refs.~\cite{SnowdenIfft:1995ke,Engel:1995gw,SnowdenIfft:1997hd,Collar:1994mj} using ancient Mica. This is due to improved read-out technology allowing for much larger exposure, improved shielding from backgrounds due to using minerals from deep boreholes instead of from close to the surface, and to using minerals different from Mica. We will show that, employing $\mathcal{O}(500)\,$Myr old minerals, paleo detectors could probe WIMP-nucleon cross sections far below current limits. For relatively light WIMPs with $m_\chi \lesssim 10\,$GeV, read-out techniques with exquisite ($\sim 1\,$nm) spatial resolution are crucial. Probing $\mathcal{O}(10)\,$mg of target material would then allow to probe SI WIMP-nucleon cross sections approximately one order above the (Xe) neutrino floor. Excellent track length resolution is less crucial for heavier WIMPs since they induce longer tracks, yet larger samples are required to compete with existing experiments. For $m_\chi \gtrsim 50\,$GeV, scattering cross sections $1-2$ orders of magnitude below current limits can be probed by reading out $\mathcal{O}(100)\,$g of target mass.

\section{Dark Matter Signal} \label{sec:Signal}
The differential recoil rate per unit target mass for a WIMP with mass $m_\chi$ scattering off a target nucleus with mass $m_T$ is given by
\begin{equation} \label{eq:diffRecoilRate}
	\left(\frac{dR}{dE_R}\right)_T = \frac{2 \rho_\chi}{m_\chi} \int d^3v \, v f({\bf v},t) \frac{d \sigma_T}{dq^2}(q^2, v) \;,
\end{equation}
where $E_R$ is the recoil energy, $\rho_\chi$ the local WIMP mass density, $f({\bf v},t)$ the WIMP velocity distribution, and $d\sigma_T/dq^2$ the differential WIMP-nucleus scattering cross section with the (squared) momentum transfer ${q^2 = 2 m_T E_R}$. For canonical  SI WIMP-nucleon couplings, the differential WIMP-nucleus cross section is well approximated by
\begin{equation} \label{eq:WIMPdiffxsec}
	\frac{d\sigma_T}{dq^2}(q^2,v) = \frac{A_T^2 \sigma_n^{\rm SI}}{4 \mu_{(\chi n)}^2 v^2} F^2(q) \theta(q_{\rm max}-q),
\end{equation}
where $A_T$ is the number of nucleons in the target nucleus, $\sigma_n^{\rm SI}$ the SI WIMP-nucleon scattering cross section at zero momentum transfer, and $\mu_{(\chi n)}$ the reduced mass of the WIMP-nucleon system. The form factor $F(q)$ accounts for the finite size of the nucleus; we use the Helm form factor\footnote{More refined calculations of the form factors are available, although only for a few isotopes, see e.g. Refs.~\cite{Vietze:2014vsa,Gazda:2016mrp,Korber:2017ery,Hoferichter:2018acd}}~\cite{Helm:1956zz, Lewin:1995rx, Duda:2006uk}. The Heaviside step function $\theta(q_{\rm max}-q)$ accounts for the maximal momentum transfer $q_{\rm max} = 2 \mu_{(\chi T)} v$ given by the scattering kinematics. For the velocity distribution we use a Maxwell-Boltzmann distribution truncated at the galactic escape velocity and boosted to the Earth's rest frame as in the Standard Halo Model~\cite{Drukier:1986tm,Lewin:1995rx,Freese:2012xd}.

\begin{figure}
   \includegraphics[width=\linewidth]{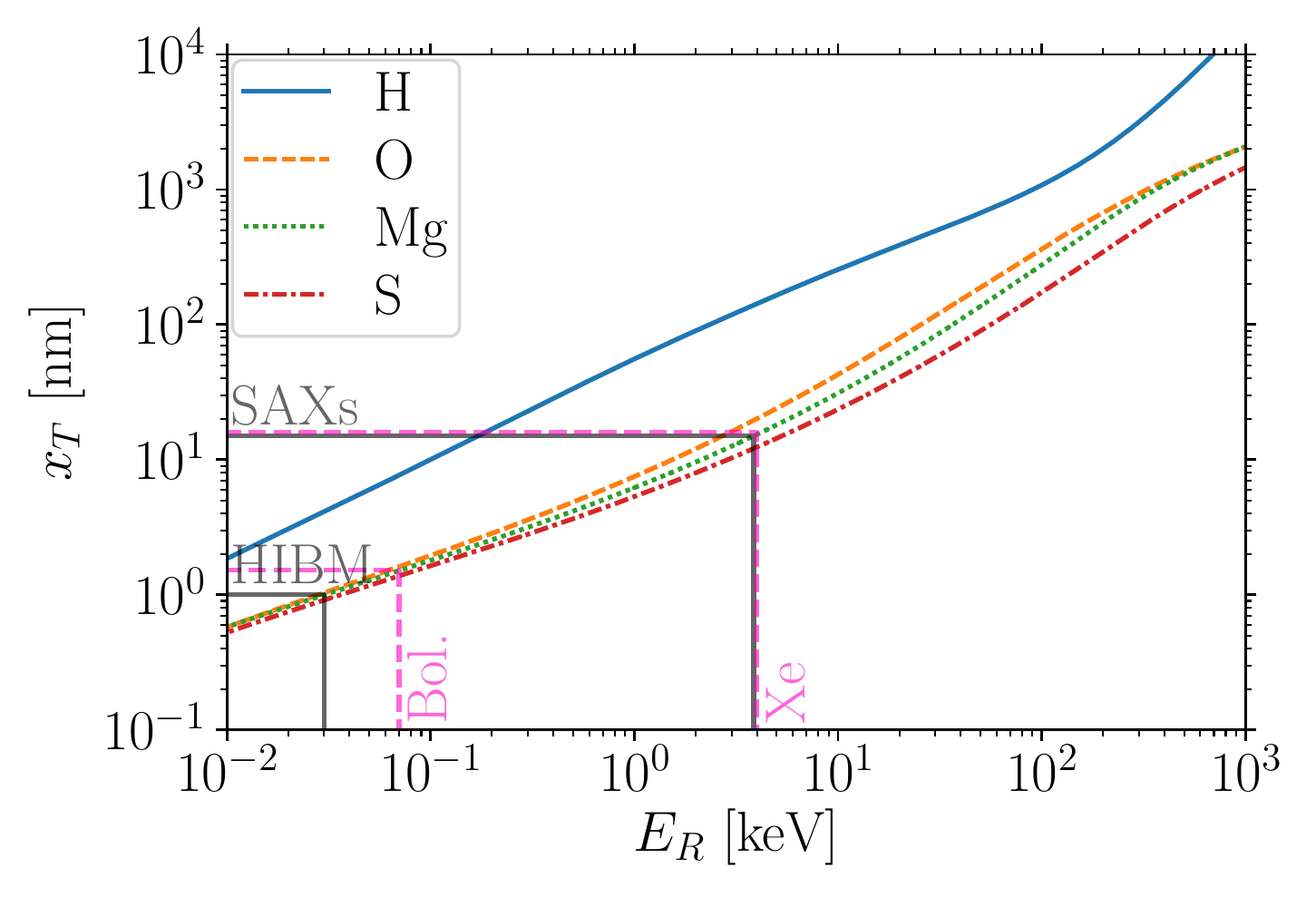}
   \caption{Range $x_T$ of ions $T$ with initial energy $E_R$ for the four different constituents of epsomite [\ccEps] as labeled in the legend. The two gray horizontal lines, with track lengths of $1\,$nm and $15\,$nm, correspond to the shortest tracks which could be read out with Helium Ion Beam Microscopy (HIBM) and Small Angle X-ray scattering (SAXs) respectively, as discussed further in Sec.~\ref{sec:ReadOut_Sens}.  The vertical gray lines indicate the corresponding recoil energy threshold. Similarly, the pink lines indicate approximate recoil energy thresholds for conventional direct detection experiments using liquid Xe time projection chamber (TPC)~\cite{Akerib:2016vxi,Aprile:2018dbl} or cryogenic bolometer (Bol.)~\cite{Angloher:2015eza,Agnese:2016cpb} detectors. One can see that the resolution of SAXs ($\sim 15\,$nm) gives a threshold comparable to liquid Xe TPC experiments.  Likewise, the resolution of HIBM ($\sim 1\,$nm) corresponds to a nuclear recoil energy threshold of some tens of eV, comparable to (or somewhat better than) what cryogenic bolometers can achieve.}
   \label{fig:Range}
\end{figure}

From the differential recoil rate for given target nuclei we compute the associated spectrum of ionization track lengths. The stopping power for a recoiling nucleus incident on an amorphous target, $dE / dx_T$, is obtained from the \texttt{SRIM} code\footnote{The results of this code differ by $\sim 10\%$ from results of analytical calculations of Ref.~\cite{Wilson:1977jzf}, which extends the Lindhard theory~\cite{Lindhard:63I,Lindhard:63II} to lower energies as required for WIMP-induced recoils.}~\cite{Ziegler:1985,Ziegler:2010} and the ionization track length for a recoiling nucleus with energy $E_R$ is
\begin{equation} \label{eq:TrackLength}
	x_T(E_R) = \int_0^{E_R} dE \left| \frac{dE}{dx_T}(E) \right|^{-1} \;.
\end{equation}
As an example, Fig.~\ref{fig:Range} shows the track length for the different elements comprising epsomite [\ccEps] as a function of recoil energy. For any particular element in a given target material, a given recoil energy $E_R$ is in one-to-one correspondence to a track length $x_T$. However, in practice it is extremely challenging to identify which element gave rise to a given track in a compound target mineral\footnote{The width and shape of a track does carry information about the ion which gave rise to a given track. Accessing this information would however require detailed resolution of the track profiles; the diameter of tracks in typical materials is a few nanometers.}. The relevant quantity is thus the span of possible recoil energies (of the different elements) corresponding to a measured track length. Note that a given track length corresponds to significantly smaller recoil energy for lighter nuclei (e.g. H) than for heavier nuclei because the stopping power increases with the charge of the nucleus. Throughout this letter, we assume that the stopping power for H and He nuclei is too small for such ions to give rise to recoverable tracks; see Ref.~\cite{Drukier:2018pdy} for a discussion of the effects if H and He tracks could be measured. We note that such figures look similar for other minerals; quantitative differences arise mainly from different molecular number density. In general, lower molecular number densities yield longer tracks for the same recoiling nucleus and recoil energy.

The finite spatial resolution of any given read-out technique further weakens the relation between measured track lengths and nuclear recoil energies. In particular for track lengths comparable to (or smaller than) the read-out resolution, the measured length can differ significantly from the true length of the track. Thus, Fig.~\ref{fig:Range} also allows to obtain the effective recoil energy threshold for paleo detectors for various read out techniques by demanding the true track length to be larger than the spatial resolution. Using Helium Ion Beam Microscopy, discussed further in Sec.~\ref{sec:ReadOut_Sens}, we expect that tracks as short as $1\,$nm could be read out, corresponding to a recoil energy threshold of $E_R \gtrsim 20\,$eV in epsomite. Small Angle X-ray scattering is expected to be able to reconstruct tracks with $x_T \gtrsim 15\,$nm, cf. Sec.~\ref{sec:ReadOut_Sens}, corresponding to a recoil energy threshold of $E_R \gtrsim 4\,$keV. For comparison, we also indicate typical recoil energy thresholds for direct detection experiments using liquid Xe time projection chamber~\cite{Akerib:2016vxi,Aprile:2018dbl} or cryogenic bolometer~\cite{Angloher:2015eza,Agnese:2016cpb} detectors. 

\begin{figure}
	\includegraphics[width=\linewidth]{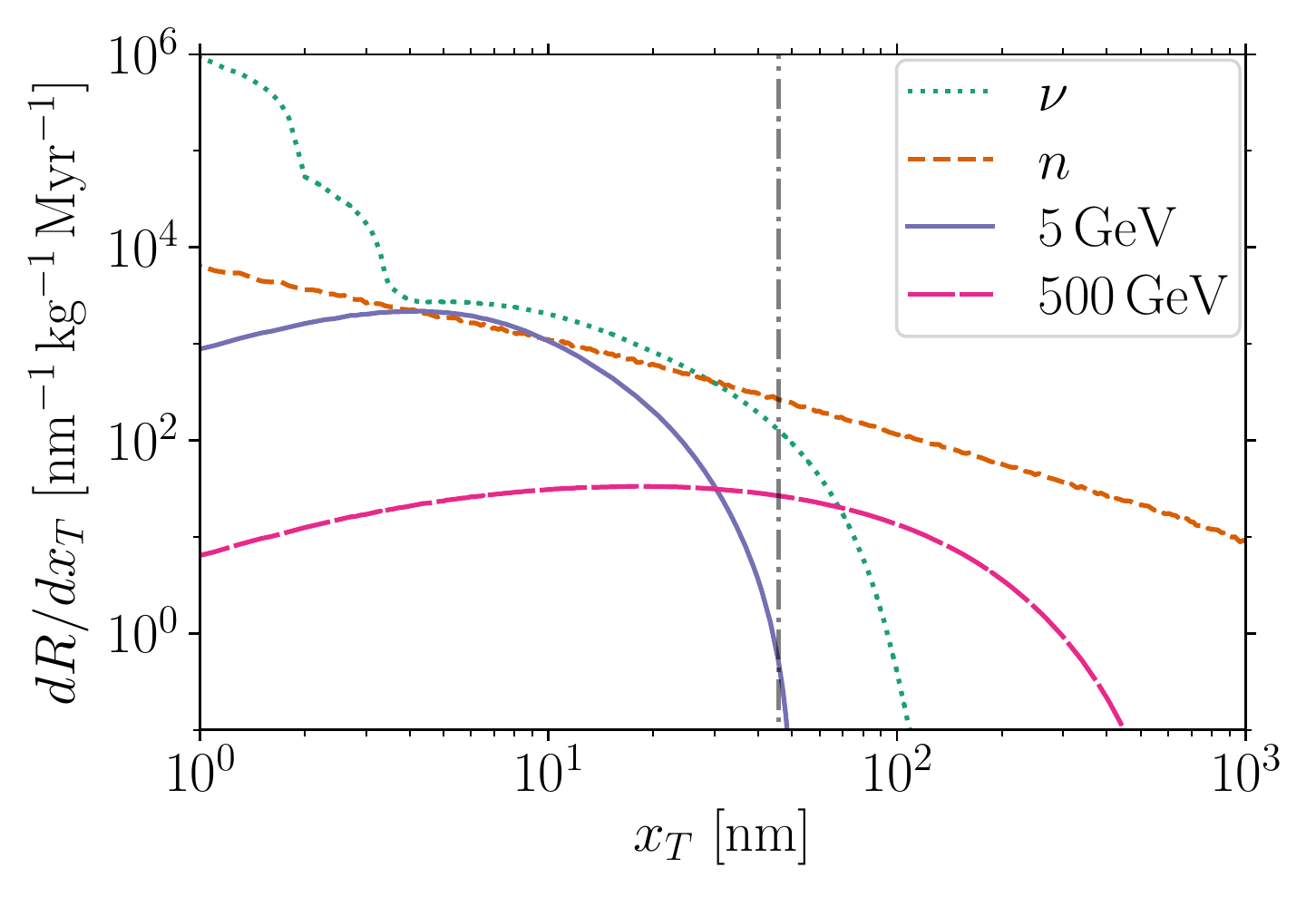}
	\caption{Track length spectra from recoiling nuclei in epsomite induced by background neutrinos ($\nu$, dotted), neutrons ($n$, short-dashed) and WIMPs with $m_\chi = 5\,$GeV (solid) and $m_\chi = 500\,$GeV (long-dashed), assuming $\sigma_n^{\rm SI} = 10^{-45}\,{\rm cm}^2$. The vertical dash-dotted line indicates the length of $^{234}$Th tracks from ($^{238}{\rm U} \to ^{234}{\rm Th} + \alpha$) decays. Resulting sensitivity projections for WIMP signals are shown in Fig.~\ref{fig:Reach}.}
	\label{fig:dRdx}
\end{figure}

Together with the recoil spectra obtained from Eq.~\eqref{eq:diffRecoilRate} and summing over the contributions from the different target nuclei in the mineral weighted by the respective mass fraction in the target, we obtain the track length spectra for nuclear recoils induced by WIMPs. In Fig.~\ref{fig:dRdx}, we plot track length spectra induced by WIMPs incident on epsomite for $m_\chi = 5\,$GeV (solid) and $m_\chi = 500\,$GeV (dashed), assuming $\sigma_n^{\rm SI} = 10^{-45}\,{\rm cm}^2$, as well as backgrounds discussed below.

\section{Backgrounds} \label{sec:Backgrounds}
There are a number of possible background sources which must be mitigated. Nuclear scattering events arising from cosmic ray induced neutrons are greatly suppressed by using minerals from deep below ground. In the deepest current laboratories at $\mathcal{O}(2)\,$km rock overburden, the muon induced neutron flux is $\mathcal{O} (10^6)\,{\rm cm}^{-2}\,{\rm Gyr}^{-1}$~\cite{Mei:2005gm}. We expect a neutron flux of $\mathcal{O}(10^2)\,{\rm cm}^{-2}\,{\rm Gyr}^{-1}$ by using minerals obtained from a depth of 5\,km; at depths of $6\,$km the expected flux is $\mathcal{O}(10)\,{\rm cm}^{-2}\,{\rm Gyr}^{-1}$.\footnote{At depths larger than $\sim 6\,$km rock overburden, neutron production from atmospheric neutrinos interacting with nuclei in the vicinity of the target must be taken into account in addition to the neutrons produced by cosmogenic muons, see e.g. Ref.~\cite{Aharmim:2009zm}. While the Earth is relatively opaque to muons produced in the atmosphere, the flux of atmospheric neutrinos is roughly constant at any depth and the interactions of such neutrinos in the Earth provide for a ``floor" in the cosmogenic neutron background, which cannot be suppressed at greater depths.} For minerals obtained from depths larger than $\sim 5\,$km of rock overburden, the background due to neutrons induced by cosmic rays is negligible compared to other background sources discussed below.

Background from intrinsic radioactivity can be suppressed by choosing target minerals (e.g. UBRs or MEs) which have low levels of contamination by highly radioactive elements such as U or Th. Out of the contaminants, $\beta$ and $\gamma$ emitters do not give rise to relevant background as they lead only to $E_R \lesssim 10\,$eV recoils of the daughter nuclei\footnote{$\beta$-decays of rare isotopes of light elements, which can e.g. be produced by spontaneous fission of heavy nuclei or in spallation processes induced by atmospheric neutrinos, can yield recoils of the daughter nuclei as large as $E_R = \mathcal{O}(100)\,$keV. Such isotopes are not produced frequently enough relative to other neutrino or neutron induced backgrounds to significantly impact our sensitivity projections.} and the emitted electrons and photons do not produce persistent tracks themselves. $\alpha$ emitters are a more serious problem. Due to the $\mathcal{O}(500)\,$Myr integration time, typically the entire decay chains of heavy radioactive elements such as U and Th are recorded in the mineral, leading to characteristic patterns of recoil tracks~\cite{SnowdenIfft:1995ke}. Such patterns allow for efficient rejection of these backgrounds.

However, any target mineral will contain a population of events where only the first $\alpha$ decay has taken place. The most problematic source of such \textit{single-$\alpha$}  (1$\alpha$) events is the ($^{238}{\rm U} \to {^{234}{\rm Th}} + \alpha$) decay in the $^{238}$U decay chain~\cite{Collar:1995aw,SnowdenIfft:1996zz}. Ideally one would be able to reconstruct the $\sim 10\,\mu$m $\alpha$-track in order to reject these events.  Even with the pessimistic assumption that the $\alpha$ tracks are not visible,  the characteristic energy $E_R = 72\,$keV of the $^{234}$Th nucleus from the $\alpha$-decay of $^{238}$U can be used to reject this background.

The most relevant background induced by radioactive sources are fast neutrons. Such neutrons arise both from spontaneous fission (SF) events as well as from $(\alpha,n)$ reactions of the $\alpha$ particles from radioactive decays with nuclei in the target volume.\footnote{Due to the range of $\mathcal{O}({\rm MeV})$ neutrons in natural minerals, background modeling in principle requires simulating neutron production and propagation in an $\mathcal{O}({\rm m}^3)$ volume surrounding the target volume. For simplicity, we approximate the target volume as being part of an ``infinitely'' large mineral of homogeneous chemical composition in our background simulation. See the Appendix of Ref.~\cite{Baum:2019fqm} for further discussion.} Note that SF events themselves are easily rejected since the two heavy recoiling nuclei with typical energies of order 100\,MeV and their subsequent decays lead to track signatures even richer than chains of alpha decays.

We calculate the neutron spectra from SF and $(\alpha,n)$ interactions with the \texttt{SOURCES} code~\cite{sources4a:1999} and the induced nuclear recoil spectrum with our own Monte Carlo (MC) simulation. Note that our MC takes into account only elastic neutron-nucleus scattering, yielding a pessimistic estimate of the induced background. Other interactions such as inelastic scattering and $(n,\alpha)$ interactions absorb a larger fraction of the neutron energy than elastic scattering. Neutron-induced backgrounds are crucial to determine which target minerals are best suited for paleo-detectors: Optimal targets contain H, which moderates the fast neutrons effectively, and do not contain Li or Be since these elements have large $(\alpha,n)$ cross sections leading to too large background.

Another relevant background arises from coherent scattering of neutrinos off nuclei in the mineral, where the neutrinos come from the Sun, supernovae, and atmospheric cosmic ray interactions~\cite{Billard:2013qya}. We use neutrino fluxes from Ref.~\cite{OHare:2016pjy}.

The nuclear recoil spectrum due to the neutrinos, neutrons, and $1\alpha$ backgrounds are converted to ionization track length spectra analogously to the WIMPs. In Fig.~\ref{fig:dRdx}, we plot the induced track length spectra in epsomite. Note, that the normalization of the neutrino- and radioactivity-induced backgrounds is fixed by the neutrino-flux and the concentration of $^{238}$U. The WIMP induced recoil spectra scale with $\sigma_n^{\rm SI}$.

\section{Sensitivity Projection} \label{sec:ReadOut_Sens}

A number of experimental techniques can potentially be used to reconstruct the damage tracks in paleo-detectors. The highest spatial resolution can be achieved with electron, atomic force, or focused ion beam microscopes. More conventional microscopes using light from X-rays to the optical band allow to image much larger quantities of material at the cost of lower spatial resolution; cf. Ref.~\cite{Drukier:2018pdy} for a broader discussion of read-out techniques. Here, we consider two benchmark scenarios in order to project the sensitivity of paleo-detectors: 
\begin{itemize}
   \item Helium Ion Beam Microscopy (HIBM)~\cite{Hill:2012} could achieve exquisite spatial resolution of $\sigma_x \approx 1\,$nm. HIBM allows to read out a $\mathcal{O}(100)\,$nm thick layer of a sample. Bulk read-out can be realized by iteratively reading out a layer with HIBM and ablating the layer away using focused (higher $Z$) ion beams~\cite{Lombardo:2012,Joens:2013} and/or pulsed lasers~\cite{Echlin:2015,Pfeifenberger:2017,Randolph:2018}. We assume that such a technique would allow for the processing of $\mathcal{O}(10)\,$mg of target mass. For $1\,$Gyr old minerals, this corresponds to an exposure of $\varepsilon = 0.01\,$kg\,Myr.
   \item Small Angle X-ray scattering (SAXs) tomography at synchrotron facilities~\cite{Schaff:2015} has achieved spatial resolution of $\sigma_x \sim 15\,$nm~\cite{Holler:2014}. It allows for three-dimensional imaging of bulk materials and we assume that samples as large as $\mathcal{O}(100)\,$g could be processed, corresponding to $\varepsilon = 100\,$kg\,Myr.
\end{itemize}
We note that the (effective) three-dimensional reconstruction of damage tracks arising from nuclear recoils has not yet been demonstrated with HIBM or high-resolution SAXs tomography. Thus, although the benchmark read-out scenarios we consider here are promising, we are proposing challenging applications of these techniques.

In the next section, we present the projected sensitivity for these two read out scenarios. In order to estimate the sensitivity, we use a simple cut-and-count analysis: First, we account for the finite track length resolution by sampling the spectrum and smearing each event
\begin{equation}
   x_T^{\rm obs} = x_T^{\rm true} + \Delta_x(\sigma_x) \;,
\end{equation} 
with the smeared (true) track length $x_T^{\rm obs}$ ($x_T^{\rm true}$) and where $\Delta_x$ is a random number with dimension length drawn from a normal distribution with standard deviation $\sigma_x$ given by the track length resolution.\footnote{In particular, the smearing turns the mono-energetic $^{234}$Th recoils from $\alpha$ events in Fig.~\ref{fig:dRdx} into a Gaussian.} 

The track length spectrum could potentially be modified by (partial) annealing of the damage tracks over geological timescales. For a discussion of annealing and the associated technique of fission track dating, we refer to Ref.~\cite{Drukier:2018pdy} and references included therein. The effects of annealing depend on the geothermal history and the precise properties of the target material, such an analysis is beyond the scope of this work. Let us note that for lower mass WIMPs with $m_\chi \lesssim 10\,$GeV, the sensitivity depends rather little on the particular target material. The dominant background for such WIMP masses arises from neutrino-induced recoils, cf. Fig.~\ref{fig:dRdx}, and thus one can use target materials which are more robust to annealing, e.g. UBRs. For heavier WIMPs, the dominant background source are radiogenic neutrons, and the sensitivity is thus much more dependent on the suppression of the radiogenic backgrounds. As discussed in Sec.~\ref{sec:Backgrounds}, MEs promise particularly good suppression of radiogenic neutrons, but, on the other hand, damage tracks potentially anneal faster than in UBRs. However, for heavier WIMPs, damage tracks are formed from more energetic recoils than for lighter WIMPs, and are thus expected to be more robust to annealing.

Additional effects which depend on the geological history of the target mineral, such as the stability of the minerals as they are buried or extracted, could also effect the sensitivity of paleo-detectors. For example, as discussed in Ref.~\cite{Gorham:2001wr}, the low density of halite causes increased plasticity at higher temperature and pressure associated with depths $\gtrsim 10 \,$km. While this leads to re-crystallization, ``reseting'' the damage tracks in the material, the increased mobility of the halite at depth can also lead to the formation of salt diapirs, which are salt structures such as domes or walls that extrude into the rock overlying the original salt bed and are thus more accessible to boreholes drilled from the surface~\cite{Trusheim:1960,Seni:1983,Jackson:1986,Koyi:1993,Vendeville:1991,Jackson:1994,Fort:2012}. Furthermore, the process of diapirism can render the minerals in new salt structures more radiopure than the minerals in the initial bed had been at greater depth. Thus, depending on the depth and temperature profiles of local geology over the relevant time scales, the various backgrounds in paleo-detectors could become more or less relevant when the target mineral is reset. As the modeling of such geothermal and geochemical processes on the measured track length spectra in paleo-detectors depends on the specific geological history of the deposit containing the target mineral, we leave such an analysis for future work.

In order to estimate the sensitivity, we optimize the lower and upper cutoff on the track length for each WIMP mass hypothesis by computing the minimal WIMP-nucleon cross section for which the signal-to-noise ratio satisfies
\begin{equation} \label{eq:SNR}
   \frac{S}{\sqrt{B_\nu + \Sigma_\nu^2 B_\nu^2 + B_n + \Sigma_n^2 B_n^2 + B_{1\alpha} + \Sigma_{1\alpha}^2 B_{1\alpha}^2}} \geq 3 \;,
\end{equation}
as a function of the track length cutoffs, with the smallest allowed cutoff given by $\sigma_x/2$. Here, $S$ ($B_i$) is the number of signal (background) events between the lower and upper track length cutoffs, and $\Sigma_i$ the relative systematic error of the respective background, $i = \{\nu, n, 1\alpha\}$. We also demand the number of signal events to be $S \geq 5$. 

The systematic uncertainty of the neutrino background is given by extrapolating the neutrino fluxes over the integration time of paleo-detectors, e.g. from varying supernova rates~\cite{2017arXiv170808248S} or cosmic ray fluxes. We account for this by choosing $\Sigma_\nu = 100\,\%$. For the radioactivity induced backgrounds, we assume a systematic uncertainty of $\Sigma_n = \Sigma_{1\alpha} = 1\,\%$.

\section{Results} \label{sec:Results}

\begin{figure*}
   \includegraphics[width=0.49\linewidth]{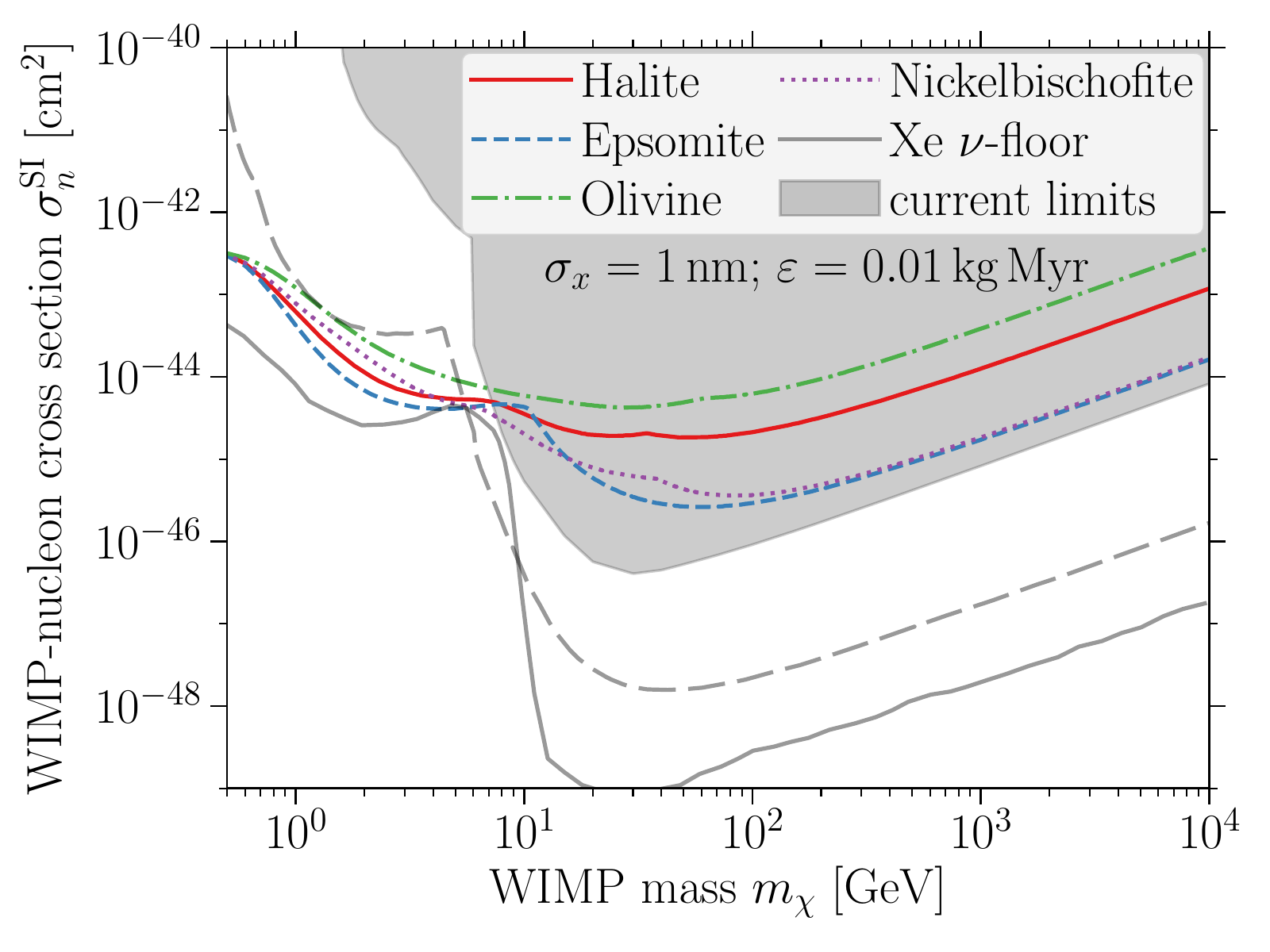}
   \includegraphics[width=0.49\linewidth]{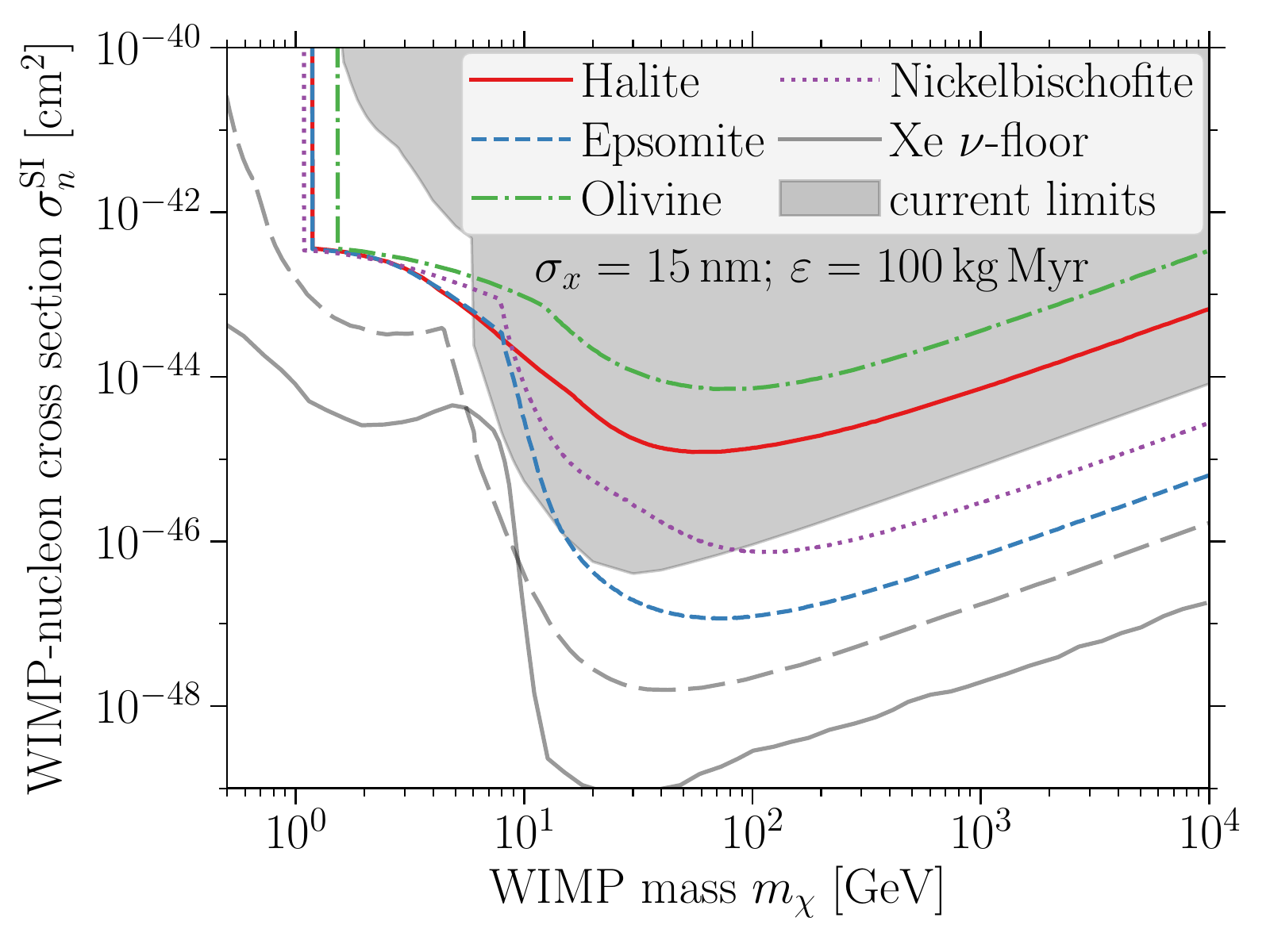}
   \caption{The left (right) panel shows the projected sensitivity to SI WIMP-nucleon scattering using HIBM (SAXs) read-out for an exposure of $\varepsilon = 0.01 (100) \,$kg\,Myr, cf. Sec.~\ref{sec:ReadOut_Sens}. The different lines correspond to different target materials as indicated in the legend. For reference, the light gray line shows the conventional neutrino-floor for direct detection experiments using Xe as target, taken from Ref.~\cite{Cushman:2013zza}. The shaded area shows current direct detection limits~\cite{Angloher:2015ewa,Agnese:2017jvy,Petricca:2017zdp,Aprile:2018dbl,Agnes:2018ves,Abdelhameed:2019hmk}. The dashed gray line indicates projections for the reach of the conventional direct detection experiments SuperCDMS~\cite{Agnese:2016cpb}, Lux-Zeplin~\cite{Akerib:2018lyp}, and XENONnT~\cite{Aprile:2015uzo} which are expected to begin taking data in the near future. Note that our sensitivity projections for paleo detectors extend to cross sections roughly an order of magnitude smaller than the cross sections these conventional experiments, in particular SuperCDMS, aim to probe at masses $m_\chi \lesssim 10\,$GeV.}
   \label{fig:Reach}
\end{figure*}

In Fig.~\ref{fig:Reach}, we show the projected sensitivity for the example minerals halite, epsomite, olivine and nickelbischofite for the HIBM (left panel) or SAXs (right panel) read-out scenarios described in Sec.~\ref{sec:ReadOut_Sens}. Comparing to current direct detection limits indicated by the shaded area in Fig.~\ref{fig:Reach}, for lighter WIMPs $m_\chi \lesssim 10\,$GeV the sensitivity of paleo-detectors is orders of magnitude better than current limits. A factor $10-100$ improvement compared to current limits is expected for heavier WIMPs $m_\chi \gtrsim 50\,$GeV. For intermediate mass WIMPs with $10\,{\rm GeV} \lesssim m_\chi \lesssim 50\,$GeV, bounds from liquid Xe experiments are comparable to our projections. We also indicate the projected sensitivity of conventional direct detection experiments expected to start taking data in the next few years in Fig.~\ref{fig:Reach}. In particular, the dashed gray line indicates projections for SuperCDMS~\cite{Agnese:2016cpb}, Lux-Zeplin~\cite{Akerib:2018lyp}, and XENONnT~\cite{Aprile:2015uzo}. For WIMPs with $m_\chi \gtrsim 10\,$GeV, the projected reach of these experiments is better than what we project here for paleo detectors. For lighter WIMPs with $m_\chi \lesssim 10\,$GeV our projections for paleo detectors extend to cross sections roughly one order of magnitude smaller than those SuperCDMS is envisaged to probe. 

This figure demonstrates the potential of paleo-detectors as well as the trade-offs between different targets and read-out techniques. The example minerals were selected as representatives of different classes of target minerals. Halite and epsomite (olivine and nickelbischofite) are MEs (UBRs) for which we assume $^{238}$U concentrations of $C^{238} = 0.01~(0.1)$\,ppb in weight; see the Appendix of Ref.~\cite{Baum:2019fqm} for a discussion of $^{238}$U concentrations in minerals. Halite and olivine are very common minerals, while epsomite and nickelbischofite have chemical compositions making them particularly well suited as target materials for paleo-detectors. In particular, they contain H which significantly reduces the neutron-induced background.

Broadly speaking, there are two background regimes: for lighter WIMPs, the background budget is dominated by neutrino induced nuclear recoils, while for heavier WIMPs the dominant background is due to radiogenic neutrons, cf. Fig.~\ref{fig:dRdx}.

Lighter WIMPs $m_\chi \lesssim 10\,$GeV give rise to relatively short tracks. Thus, excellent track length resolution is crucial. From the left panel of Fig.~\ref{fig:Reach} we find that the sensitivity to lighter WIMPs is approximately one order of magnitude worse than the neutrino floor for Xe experiments. If we would assume $\Sigma_\nu \sim 5\,\%$, as usual in the calculation of the solar neutrino induced background for direct detection experiments~\cite{Billard:2013qya,OHare:2016pjy,Cushman:2013zza}, the sensitivity of paleo-detectors would extend down to the Xe neutrino-floor. Note that differences in sensitivity between different target materials are relatively small for $m_\chi \lesssim 10\,$GeV.

For heavier WIMPs $m_\chi \gtrsim 10$\,GeV excellent track length resolution is less important since tracks are typically sufficiently long. Instead, exposure becomes more relevant. From the right panel of Fig.~\ref{fig:Reach} we find that epsomite and nickelbischofite, the targets containing H, allow for sensitivity $10-100$-fold better than current limits. For such WIMP masses, the background is dominated by neutron-induced recoils for the assumed $^{238}$U concentrations, and H is an efficient moderator of fast neutrons, significantly reducing the number of induced nuclear recoils. Further, the sensitivity of MEs is better than that of UBRs since the neutron-induced background is proportional to the $^{238}$U concentration.

Note, that apart from the finite track length resolution we assumed perfect track reconstruction. To the best of our knowledge, no reliable estimate of corrections exists for the materials considered;\footnote{The particular case of track reconstruction by cleaving and etching Muscovite Mica has been discussed in Ref.~\cite{Collar:1994mj}.} we leave such investigations to future work. Global corrections can be mitigated easily by larger exposures. The effect of energy- or $Z$-dependent corrections can only be assessed on a case-to-case basis.

\section{Discussion} \label{sec:Discussion}
We have discussed a method  for DM searches radically different from the usual Direct Detection approach. Instead of searching for WIMP-nucleus scattering in a real-time detector, we propose to examine ancient minerals for traces of WIMP interactions. Such a search integrates over geological time scales $\mathcal{O}(500)\,$Myr, yielding enormous exposure allowing for sensitivities far beyond current upper limits even for a simple cut-and-count analysis. 

While we only studied the sensitivity for canonical spin-independent WIMP-nucleus scattering in this letter, the proposed method is also sensitive to other types of WIMP-nucleus interactions, such as canonical spin-dependent interactions or more general interactions, e.g. those found in the non-relativistic effective field theory approach to WIMP-nucleon scattering~\cite{Fan:2010gt,Fitzpatrick:2012ix}. Qualitatively, we expect improvements in sensitivity with respect to current bounds similar to the spin-independent case.  Further, since multiple target nuclei would be present within a single mineral, one could search for  $A_T^2$ or other dependence, and thus determine the type of WIMP-nucleon interaction.  
 
There are a number of interesting possibilities arising from this approach we leave for future work. For example, one could use a series of crystals of different ages. The age of the oldest available crystals is larger than the period of the Sun's rotation around the galactic center. Using standard methods, the age of the crystals can be determined with an accuracy of a few \%~\cite{RENNE2001539,Wojotowitz:2003}. If the DM in the Milky Way is not smoothly distributed but instead has significant substructure~\cite{Berezinsky:2007qu,Ricotti:2009bs,Stiff:2001dq,Freese:2003tt,Zemp:2008gw}, the Earth may well have crossed such denser DM regions during the exposure time. Hence, while substructure usually renders typical direct detection experiments less sensitive due to a decrease in the local DM density, the signal one expects in ancient minerals may well increase by orders of magnitude. 

Another possibility is the use of paleo-detectors as neutrino detectors. The large exposure would allow to study the neutrino-flux from e.g. the Sun, supernovae, or cosmic rays more precisely. 

\begin{acknowledgments}
The authors thank J.~Beacom for thoughtful comments and extensive discussion of the manuscript. The authors thank J.~Blum, J.~Conrad, D.~Eichler, R.~Ewing, A.~Ferella, A.~Goobar, S.~Nussinov,  D.~Snowden-Ifft, L.~Stodolsky, H.~Sun, and K.~Sun for useful discussions.
SB, AKD, KF, and PS acknowledge support by the Vetenskapsr\r{a}det (Swedish Research Council) through contract No. 638-2013-8993 and the Oskar Klein Centre for Cosmoparticle Physics. 
SB, KF and PS acknowledge support from DoE grant DE-SC007859 and the LCTP at the University of Michigan. 
KF is grateful for funding via the Jeff and Gail Kodosky Endowed Chair in Physics.
\end{acknowledgments}

\bibliography{DMbib}

\end{document}